# The Andravida, Greece EQ (8/06/2008, Ms=7.0R). An "a posteriori" analysis for the determination of its location, occurrence time and magnitude parameters in terms of short-term predictability.

Thanassoulas[1], C., Klentos[2], V., Verveniotis, G.[3], Zymaris, N.[4]

1. Retired from the Institute for Geology and Mineral Exploration (IGME), Geophysical Department, Athens, Greece.
   e-mail: thandin@otenet.gr - URL: www.earthquakeprediction.gr

2. Athens Water Supply & Sewerage Company (EYDAP),
   e-mail: klenvas@mycosmos.gr - URL: www.earthquakeprediction.gr

3. Ass. Director, Physics Teacher at 2nd Senior High School of Pyrgos, Greece.
   e-mail: gver36@otenet.gr - URL: www.earthquakeprediction.gr

4. Retired, Electronic Engineer.

**Abstract**

The Andravida EQ, Greece (8/6/2008, Ms=7.0R) seismic parameters: location, time of occurrence and magnitude were determined "a posteriori" in an attempt to verify the predictability of the large EQs. The Earth's electric field, after its processing by de-noising techniques, was used, in conjunction with appropriate physical models, for the determination of the epicenter by triangulation. The time of occurrence was determined in very short-term mode by the use of the tidal waves **(M1, K1)** and the "strange attractor like" seismic electric precursor, while its magnitude was calculated by the application of the "lithospheric seismic energy flow model" applied on the past seismicity of the Andravida EQ regional seismogenic area. The quite accurate obtained results corroborate the validity of the methodology and suggest its use as a valuable tool for predicting large earthquakes.

## 1. Introduction.

On 8th of June, 2008 a strong earthquake occurred at the north-western part of Peloponnesus, Greece. The magnitude of this EQ was calculated by the National Observatory of Athens **(NOA)** as **Ms = 7.0** of the Richter scale. This seismic event devastated a number of villages, there were two deaths and a large number of buildings collapsed or suffered large damages.

This earthquake re-initiated in Greece the intense debate, in the scientific community, that concerns the possibility of predicting (in short-term) a large earthquake. Various scientific works were publicized through the media claiming that this EQ had been predicted in one or another way, thus creating a great deal of confusion to people without the necessary scientific knowledge.

In this work are presented the detailed seismic precursory electric signals that preceded the Andravida, Greece EQ, (8/06/2008, **Ms = 7.0R, NOA**) as were monitored by the (currently in operation monitoring network) **PYR, ATH** and **HIO** monitoring sites of the Earth's electric field. The registered data were processed according to the methodologies presented by Thanassoulas (2007). Thus, the epicentre area, the time of occurrence within a narrow (a couple of days) time window and its magnitude were subsequently calculated.

The epicentre location of the Andravida EQ is shown as blue concentric circles in figure **(1)**. The background of the map is the seismic potential of the Greek territory, calculated for the year 2000 (see attached scale) while the thick grey lines represent the deep lithospheric fracture zones – faults (Thanassoulas, 1998, 2007).

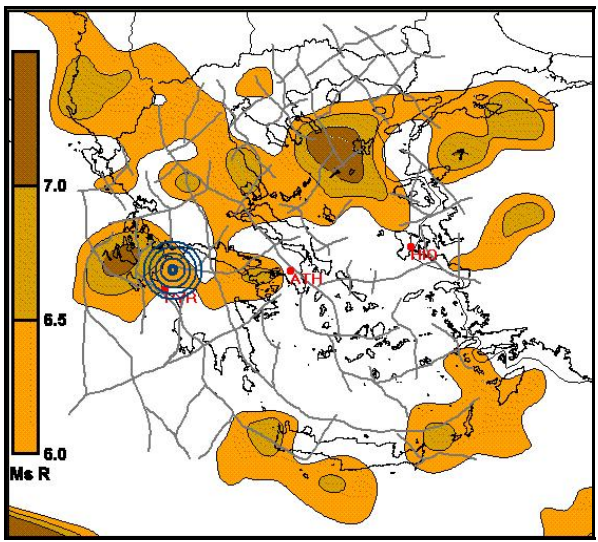

Fig. 1. Location of the Andravida EQ epicenter (blue circles) on top of the seismic potential map of Greece (calculated for the year 2000) and the deep lithospheric fracture zones–faults (thick grey lines). The minimum threshold of the seismic potential corresponds to an EQ of **Ms = 6.0 R**. Monitoring sites of **PYR**, **ATH** and **HIO** are denoted by red capital letters.

## 2. Data presentation and analysis.

Three data sets are used in this work. The first data set is the Earth's electric field registered by the **PYR, ATH** and **HIO** monitoring sites in Greece. These data are firstly used for the determination of the epicentral area of the Andravida EQ, secondly for an early warning for a pending earthquake due to the presence of seismic precursory electric signals and finally for narrowing down the estimated time window of the EQ occurrence time. These data can be downloaded from the www.earthquakeprediction.gr internet address. The second data set is the tidal variation of the Earth's gravity field. This data set has been calculated by the Rudman et al. (1977) method. It is well known that tidal waves can trigger large EQs at seismogenic areas being under exceptional stress load (Thanassoulas, 2007). The tidal variation "suggests" favorable timing window when a large EQ can take place provided that there exist the adequate stress load in the seismogenic area. Finally, the earthquake catalog of National Observatory of Athens **(NOA)** is used for the determination of the cumulative seismic energy release of the Andravida seismogenic area and of the corresponding expected maximum magnitude of the pending EQ. The EQs catalog that spans from 1901 to 2010 can be downloaded from the www.earthquakeprediction.gr internet address.

### 2.1. Earth's electric field data.

The Earth's electric field which will be analyzed spans from February 10th, 2008 to June 10th, 2008. Actually, it is four months recording that preceded the Andravida EQ. The recorded Earth's electric field has been normalized to equal length of **N-S** and **E-W** dipoles and to **N-S / E-W** directions. In the following figure **(2)** the normalized recorded electric field by the **PYR** monitoring site is presented. The considered seismic precursory electric signal is indicated by a red ellipse.

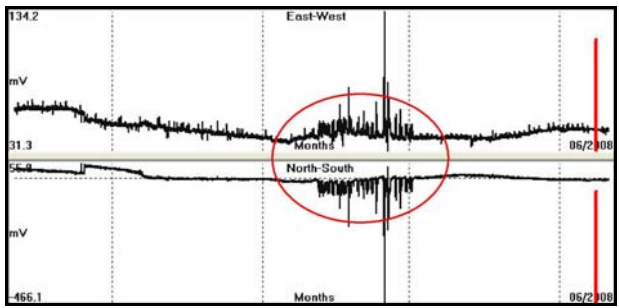

Fig. 2. **EW** and **NS** components (normalized values) of the Earth's electric field (grad) recorded by **PYR** monitoring site during the period: 10th February - 10th June of 2008. The red bar indicates the time of occurrence of the large EQ of Andravida (**Ms = 7.0R**). Note the "seismic precursory signals" (in red ellipse) recorded in both components during April.

The normalized Earth's electric field recorded by the **ATH** monitoring site is presented in figure **(3)**. Well before the occurrence of the Andravida EQ (red bar), a long sequence of **SES** electric pulses is observed. Moreover, short before the Andravida EQ, a large amplitude and short-period electric pulse is noticed.

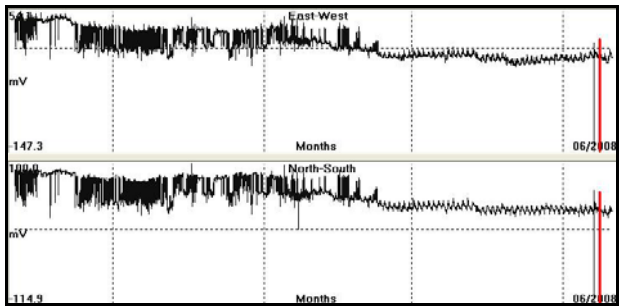

Fig. 3. **EW** and **NS** components (normalized values) of the Earth's electric field (grad) recorded by **ATH** monitoring site during the period: 10th February - 10th June of 2008. The red bar indicates the time of occurrence of large Andravida EQs (**Ms = 7.0R**). Note the "electric spike" that just preceded the Andravida EQ.

This pulse is shown in the following figure **(4)** magnified in a shorter time window of seven days from June 4th, to June 10th. Actually this signal did occur one day before the Andravida EQ occurrence. Its presence at one day before the large seismic event is important in terms of short-term prediction as it will be explained later on.

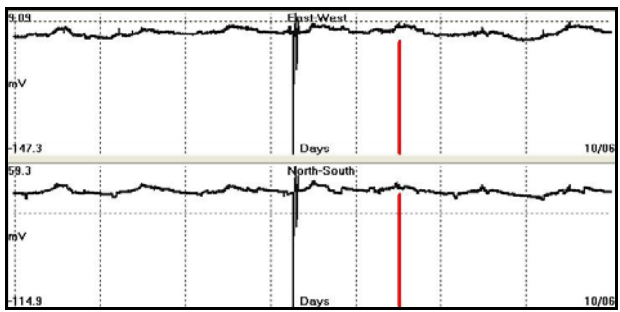

Fig. 4. The observed electric spike prior to the occurrence of the Andravida EQ (red bar). Recording period: 4th June - 10th June 2008. Note the seismic electric precursor occurrence at one day before the main seismic event.

The normalized Earth's electric field recorded by the **HIO** monitoring site is presented in figure **(5).** Apart from a very long wavelength anomaly of the recorded Earth's electric field that lasts almost for three months, no other significant seismic precursor is observed.

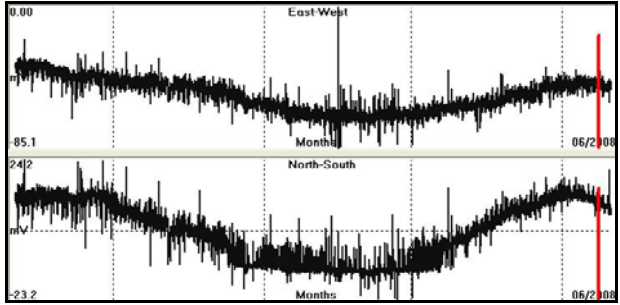

Fig. 5. **EW** and **NS** components (normalized values) of the Earth's electric field (grad) recorded by **HIO** monitoring site during the period: 10[th] February - 10[th] June of 2008. The red bar indicates the time of occurrence of large Andravida EQs (**Ms = 7.0R**).

## 2.2. Determining the epicentral area of the Andravida EQ.

The registered Earth's electric field data, from all three monitoring sites, were checked as far as it concerns the calculation of the azimuthal direction of the intensity vector of the recorded electric field at each monitoring site in relation to the actual epicenter of the Andravida EQ. Although in all cases "anomalous" electric field had been recorded prior to the occurrence of the Andravida seismic event, only **PYR** and **HIO** monitoring sites provided a stable acceptable solution. The latter is attributed probably to the short (20 meters) electric dipoles used in **ATH** monitoring site in contrast to the larger ones (160 ÷ 200 meters) used in **PYR** and **HIO** monitoring sites. The used short length dipoles in **ATH** practically behave as high-pass filters thus allowing short electric pulses (the ones shown in figure **3**) to be recorded and longer period electric signals to be rejected at all. Consequently, the epicenter of the Andravida EQ was determined only from **PYR** and **HIO** Earth's electric field recordings.

The normalized data of the Earth's electric field was "injected" by noise (Thanassoulas et al. 2008) in order to obtain the true seismic precursory signal that preceded the Andravida EQ. The results of this procedure that was applied on the data of **PYR** monitoring site are presented in the following figure **(6)**.

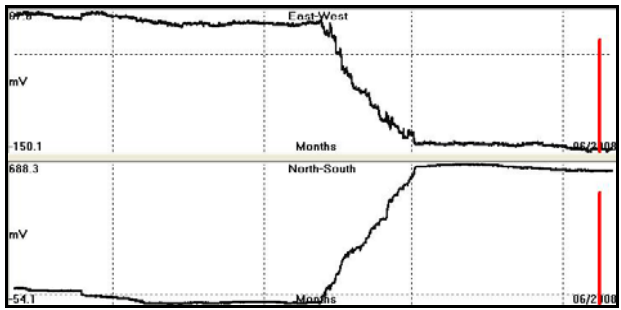

Fig. 6. Processed data of **PYR** monitoring site by the "noise injection" method. Recording period: 10[th] February - 10[th] June 2008. Noise injection parameter **p = 0.625.**

The seismic precursory signal (increased amplitude electric spikes) that is shown in figure **(2)** has been transformed in to a gradually increasing step of the electric field that corresponds to the same period of time. Next, the azimuthal direction of the Earth's electric field intensity vector was calculated at minute's samples for all the recording period (172.800 data points). The obtained average value of the azimuthal direction is: **1.75 rad or 98 degrees**. Graphically this operation is presented in the following figure **(7).** The red color of the graph shows the azimuthal results of all data points while the green line shows the average value.

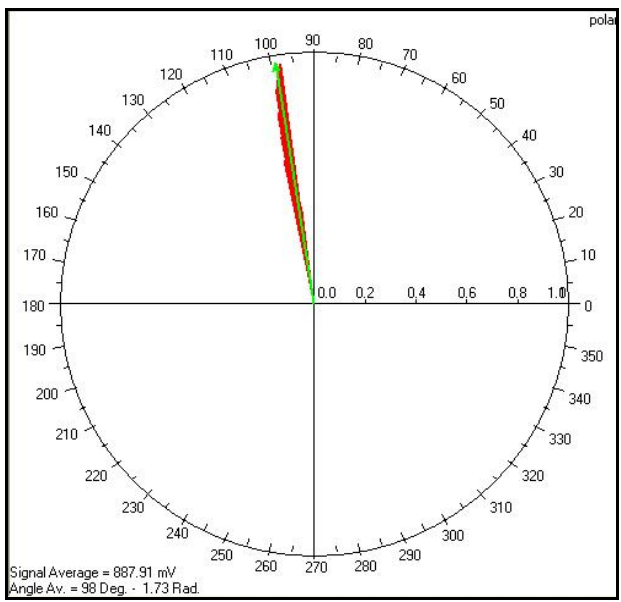

Fig. 7. Earth's electric field intensity vector that corresponds to the processed **PYR** data. The angle counts from east **(E, zero angle)** anticlockwise. Green line = average value. Red lines = individual azimuthal calculations at each data point.

The very same methodology was applied on the data of **HIO** monitoring site. The results of the "noise injection" technique are shown in figure **(8).**

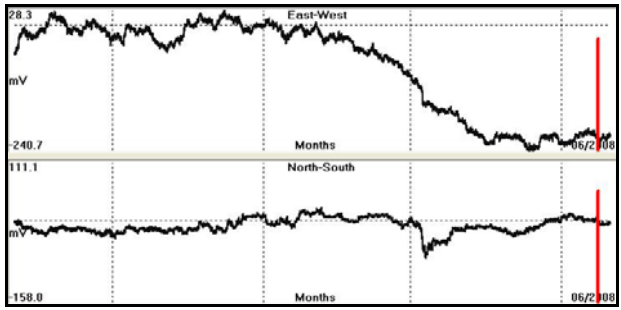

Fig. 8. Processed data of **HIO** monitoring site. Recording period: 10$^{th}$ February - 10$^{th}$ June 2008. Noise injection parameter **p = 0.625**.

It is evident in figure **(8)** that only the **E-W** component shows a seismic precursory electric signal. Moreover, that signal starts to evolve at almost start of April as the signal observed at **PYR** monitoring site did. The **N-S** component is almost flat and slightly noisy. The results from the application of the noise injection technique and the azimuthal calculation are shown in the following figure **(9)**. The obtained average value of the azimuthal direction is: **3.22 rad or 184 degrees**. The red color of the graph shows the azimuthal results of all data points while the green line shows the average value.

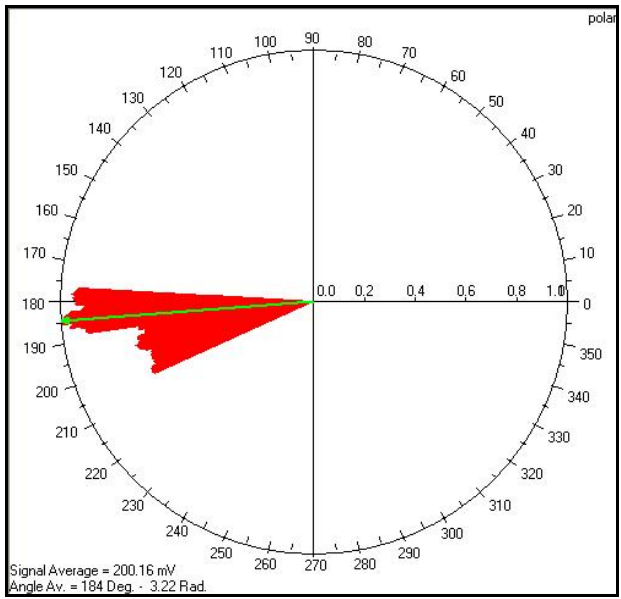

Fig. 9. Earth's electric field intensity vector that corresponds to the processed **HIO** data. The angle counts from east **(E, zero angle)** anticlockwise. Green line = average value. Red lines = individual azimuthal calculations at each data point.

At this point, having obtained two azimuthal directions from two different in location monitoring sites, it is possible by triangulation to determine the location of the electric current source and consequently, the epicentre area of the pending EQ. It is assumed that the electric current source that gives rise to the recorded electric field is located at the future focal area of the expected EQ (Thanassoulas, 1991). The azimuthal directions of figure **(7)** and **(9)** are presented in figure **(10)** simultaneously on the potential map of Greece along with the epicentre area of the Andravida EQ as it was calculated by seismological methods by **NOA** (National Observatory of Athens).

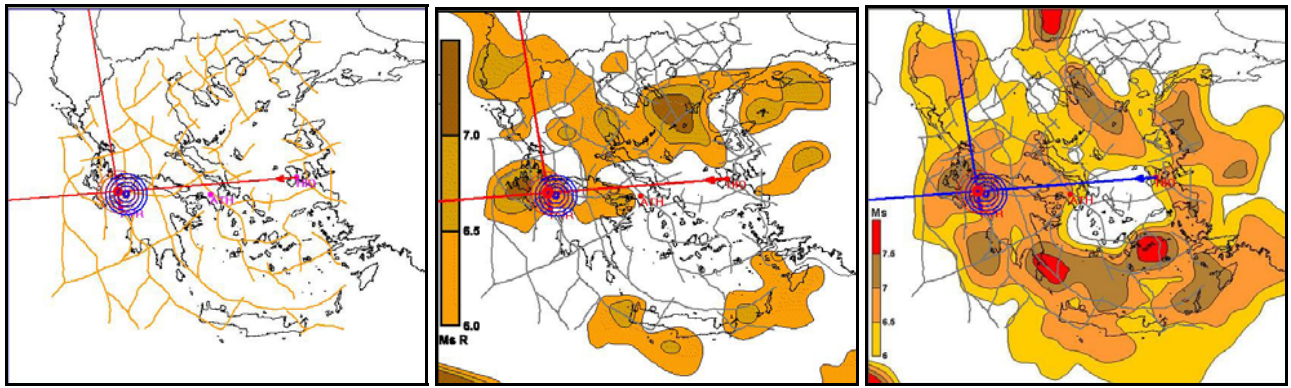

Fig. 10. Determined Andravida EQ epicentre by the analysis of the Earth's electric field compared to the seismological one. The red circles indicate the determined epicentral area. The blue circles indicate the actual location of Andravida EQ made by seismological methods. Left: white background map for easing the comparison. Middle: potential map of year 2000. Right: potential map of year 2005. Red and blue lines are the calculated intensity vectors of the electric field at the monitoring sites.

The two used different background seismic potential maps (2000, 2005) serve the purpose to demonstrate the increase of the seismic potential in Greece within only five years period of time (Thanassoulas et al. 2010). The discrepancy between the seismological epicentre and the calculated one by the Earth's electric field is of the order of 20 – 25 Km.

A great advantage for the determination of the epicentral area of a pending large EQ by the application of the latter methodology is that there are not prerequisites such as of a selectivity map so that "an estimation of the area to suffer a main shock can be obtained on the basis of the so-called selectivity map". Such a selectivity map is shown in the following figure **(11)** as it was presented by Sarlis et al. (2008). The pending EQs, after the initiation of the recorded SES, are generally expected to occur within the parallelogram.

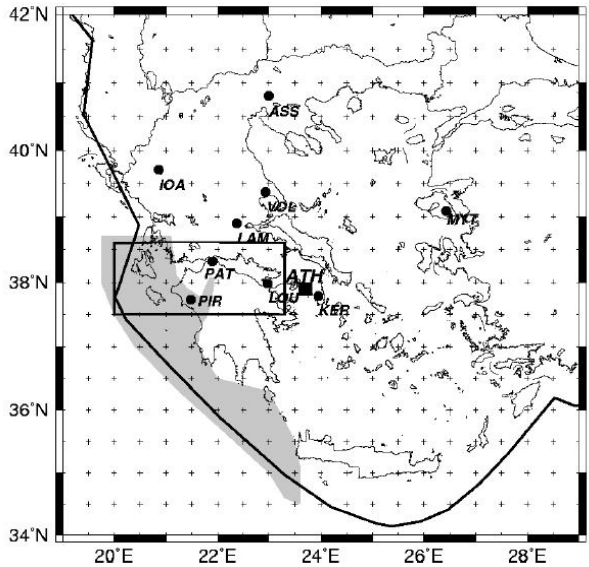

Fig. 11. Estimated area of interest (parallelogram where future EQs are expected to occur), suggested by the study of SES signals recorded by the **VAN** team (after Sarlis et al. 2008). The specific EQ of Andravida did occur (8/6/2008) within the specified parallelogram area.

It is evident that the azimuthal analysis of the Earth's electric field, as it is recorded by two or more monitoring sites, can determine (see also Thanassoulas et al. 2009) quite accurately the epicentre of a large pending EQ.

### 2.3. Time determination.

The time of occurrence of the Andravida EQ will be tested against the oscillating lithospheric plate model (Thanassoulas, 2007). Following this model, large EQs are triggered at the amplitude peaks of the lithospheric plate oscillations which are driven by the corresponding tidal waves. Consequently, the time of occurrence of the Andravida EQ is firstly compared to the yearly tidal lithospheric oscillation shown in the following figure **(12)**.

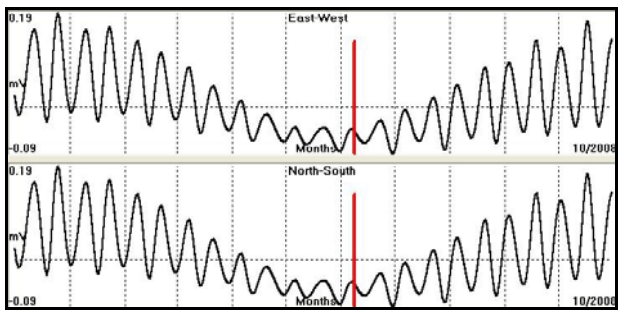

Fig. 12. Yearly (1/12/07 ÷ 30/10/2008) lithospheric oscillation (black line) superimposed by a **14** days period oscillation. The red line in the middle of the graph corresponds to the Andravida EQ.

It is clear from figure **(12)** that there is a very good agreement between the yearly tidal lithosperic oscillation maximum peak and the time of occurrence of the Andravida EQ. The latter is in agreement to the results already presented by Thanasoulas et al. (2009a) regarding the lithosperic deformation due to the yearly tidal oscillation and to the time of perihelion – aphelion of the Earth's orbit around the Sun. Practically, figure **(12)** verifies that within a years period it is highly probable, provided that the seismogenic area under study is seismically charged, that a large earthquake will be triggered approximately either on the Aphelion or on the Perihelion. Therefore, it cannot be considered as a short-term earthquake prediction.

A closer inspection of figure **(12)** shows that the Andravida EQ did occur close to a **M1** tidal peak of **T=14** days. A detailed presentation of a shorter period of time, some days before and after the Adravida EQ, is presented in the following figure **(13)**.

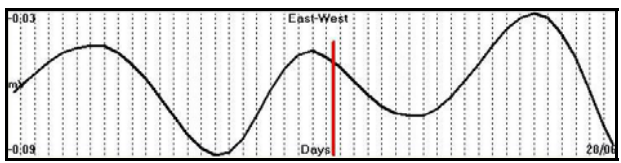

Fig. 13. **M1** lithospheric tidal oscillation (black line, **T = 14 days**) in relation to the time of occurrence of Andravida EQ (red bar).

The deviation of the occurrence time of the Andravida EQ to the corresponding **M1** tidal peak is only **1.5** days. The latter complies with similar results obtained by the analysis of a large number of EQs of the Greek territory in relation to the **M1** (**T=14** days) tidal component (Thanassoulas, 2007).

At this point it is worth to go back to figure **(4)** and to comment the presence of the electric spike that preceded the Andravida EQ by one day. The seismogenic area of the Andravida EQ, being at its final stress load phase, started to "crack" when the lithospheric plate was at its maximum deformation (peak of tidal oscillation). The initiation of the fracturing of the seismogenic area generated the electric spike that is shown in figure **(14)** while the main seismic event followed a short time (a day) after.

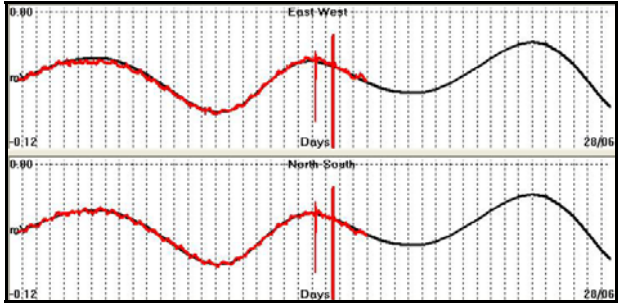

Fig. 14. Electric seismic precursor (spike) that preceded the Andravida EQ (red bar) for almost a day's time on top of the **M1 (T=14 days)** tidal oscillation (black line).

Although the present spike is the only one before the timing of the Andravida EQ and one could consider it as a precursory signal, it still cannot answer the question of: when the expected EQ will take place. Definitely a seismogenic area has been seismically charged close to fracture level and the tidal **M1** oscillation suggests some successive (1 – 2) tidal peaks when the EQ may take place. Even more, it can be speculated that the pending EQ will take place more specifically, at some day's tidal peak within a couple of days before and after the **M1** tidal peak (Thanassoulas et al. 2010a). The latter is shown in the following figure **(15).**

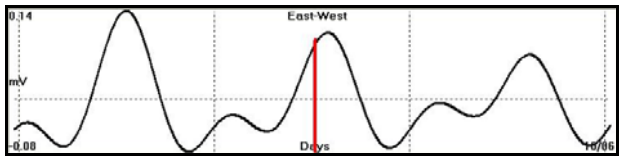

Fig. 15. **K1** lithospheric tidal oscillation (black line, **T = 1 day**) in relation to the time of occurrence of Andravida EQ (red bar). A shorter period (**T = 12 hours**) tidal component is still visible.

Actually, the Andravida EQ did occur **90** minutes before the same day's tidal peak. Even shorter deviations, than that of Andravida EQ, have been observed and reported for Skyros EQ **(26/7/2001, Ms = 6.1 R)** at **41** minutes and Kythira EQ **(1/1/ 2006, Ms = 6.9R)** at **43** minutes (Thanassoulas et al. 2010a) while more similar examples have been presented by Thanassoulas (2007).

Although it is clear that the pending EQ will take place quite close at some **M1** and **K1** tidal peak it remains yet to identify specifically the corresponding tidal peaks. This is demonstrated in the following figure **(16)**. The left part of figure **(16)** represents the tidally depended solutions regarding the timing of a large EQ (inner grey space) selected from the infinite solutions (outer white circle) valid for a rather chaotic adopted (unpredictable) seismic model. The right part of figure **(16)** represents (small inner ellipse) the even fewer solutions selected from the tidal ones, through the use of seismic electric precursors.

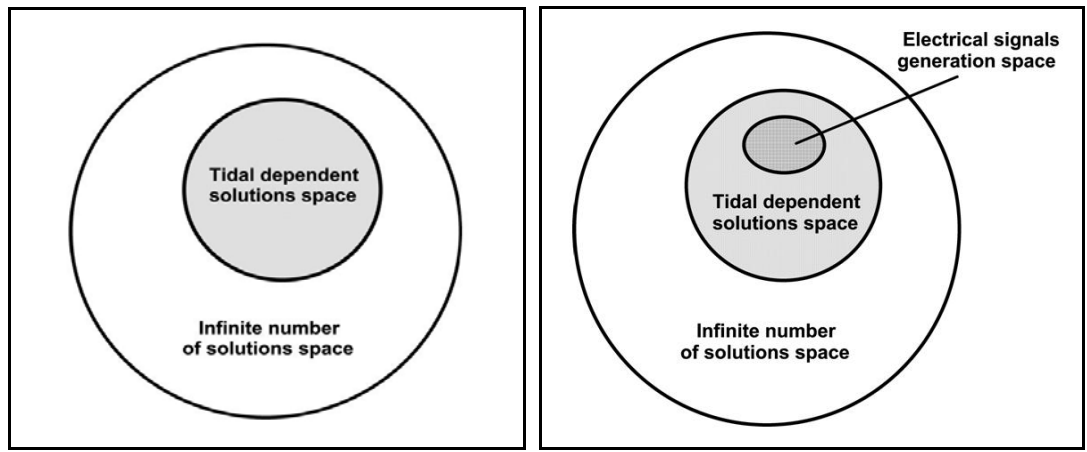

Fig. 16. The progressive shortening of the predictive time window of a large EQ is schematically presented in terms of Boolean algebra. Left = only tidal waves are taken into account. Right = further shortening of the time window by the use of electric seismic precursors (after Thanassoulas, 2007).

In the case of Andravida EQ there were various **(fig. 2, 3, 4, 5)** seismic precursory electric signals which indicated the activation of the last phase seismic charge of the seismogenic area. These precursory signals did not provided information about any shorter, than of a month, time window. The latter can be further improved by the use of the "strange attractor like" seismic electric precursor (Thanassoulas 2007, Thanassoulas et al. 2008a).

## 2.4. Application of the "strange attractor like" seismic electric precursor.

A few days before the occurrence of a large EQ, the generated at the focal area precursory oscillating electric field, when combined as registered by two distant monitoring sites, forms ellipses which usually disappear one or two days before the EQ occurrence. The latter was observed in the case of the Andravida EQ too. In the following figures **(17, 18, 19)** is presented the evolution of the "strange attractor like" for the case of Andravida EQ for the period of time from May 11$^{th}$, 2008 to June 9$^{th}$, 2008.

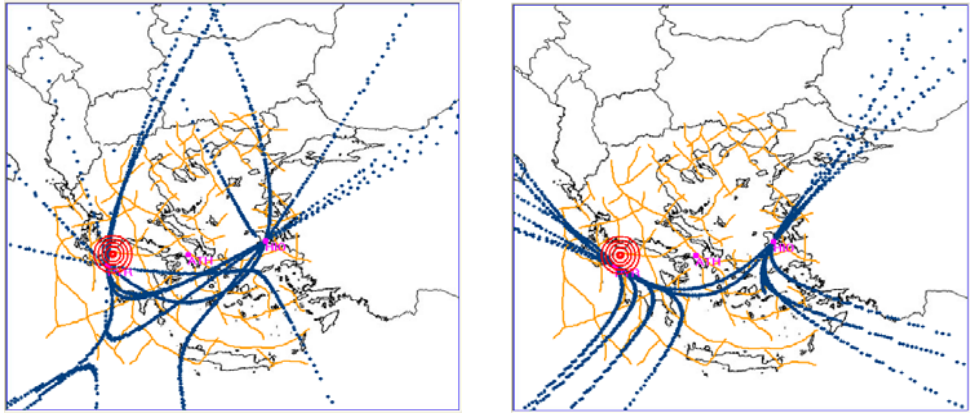

Fig. 17. "Strange attractor like" (blue dotted lines) seismic precursor observed for: left = 080511-080513 and right = 080520 – 080522 (YYMMDD mode). The absence of ellipses indicates that the seismogenic area has not reached its critical final stress load phase.

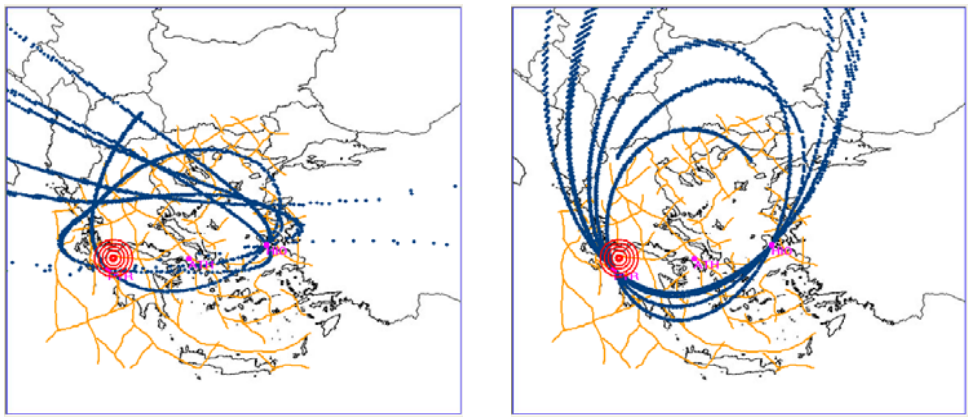

Fig. 18. "Strange attractor like" (blue dotted lines) seismic precursor observed for: left = 080529 – 080531 and right = 080601 – 080603 (YYMMDD mode). The presence of ellipses indicates that the seismogenic area has reached its critical final stress load phase.

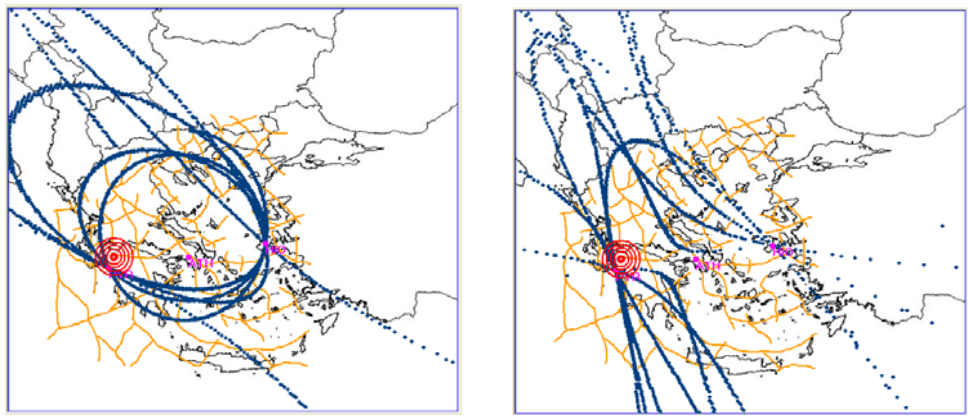

Fig. 19. "Strange attractor like" (blue dotted lines) seismic precursor observed for: left = 080604 – 080606 and right = 080607 – 080609 (YYMMDD mode). The presence of ellipses at left indicates that the seismogenic area has reached its critical final stress load phase while at right indicates that the EQ will take place at any time now since the ellipses disappeared.

Actually, the "strange attractor like" precursor disappeared on June 7$^{th}$ of 2008. Consequently, this fact specifies a very short time window of a couple of days (7÷8/6/2008) within which the pending EQ will take place. Starting from this hint and going back to the tidal components of **M1 (T=14 days)** and **K1 (T=24hours)** it is possible to identify the candidate **M1 (6÷7/6/2008)** and **K1** tidal peaks when the pending EQ will probably take place. For the case of **M1** tidal component it is shown in figure **(13)** that the deviation of the occurrence time is only **1.5 days**. For the case of **K1** tidal component it is shown in figure **(15)** that the deviation is only **90 minutes**. Therefore, if a prediction were made by using these data and a time window was accepted of two days regarding **M1** and two hours regarding **K1** (which is practically an immediate prediction) then that prediction could be accepted as a successful one.

### 2.5. Magnitude determination.

The next seismic parameter to be determined, for completing the Andravida EQ prediction, is the calculation of its magnitude. To this end the seismic energy release over past time is manipulated following the "lithospheric seismic energy flow model" introduced by Thanassoulas (2008). Before this analysis starts it is interesting to obtain a general picture of the seismic status of the Greek territory considered as a unit seismogenic area as it was before the Andravida EQ as follows.

#### 2.5.1. Greece as a unit regional seismogenic area.

A common parameter in time which can be studied for a seismogenic area is its seismic energy release. The cumulative seismic energy release for the time period of 1901 to 2008 for the Greek territory (see figure **1**) is shown in the following figure **(20)**. The cumulative seismic energy release has been calculated at a sampling interval of one month.

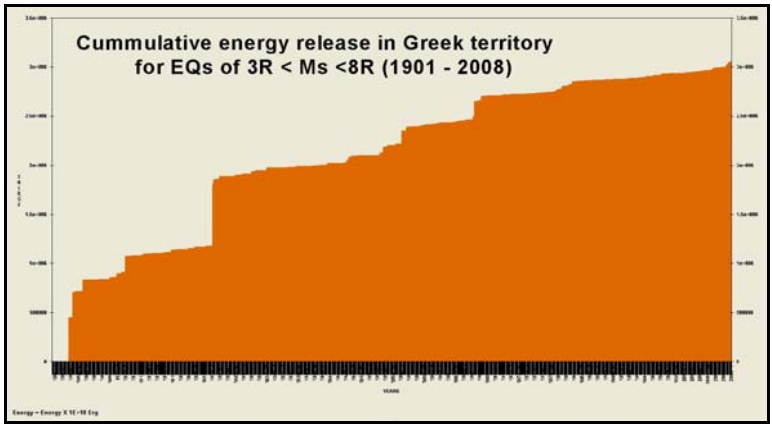

Fig. 20. Cumulative seismic energy release is shown of the Greek territory for the time period of 1901 to 2008. Sampling interval = 1 month.

It is very interesting to notice that the energy release has decreased during the last **20** years which in turn indicates excess seismic energy store in the Greek territory. Moreover, at the end of this study period a slight indication is present of an activated accelerating deformation. Although seismic energy is stored, for almost the same period of time the number of EQs has increased drastically, regardless their magnitude. The latter is shown in the following figure **(21)**.

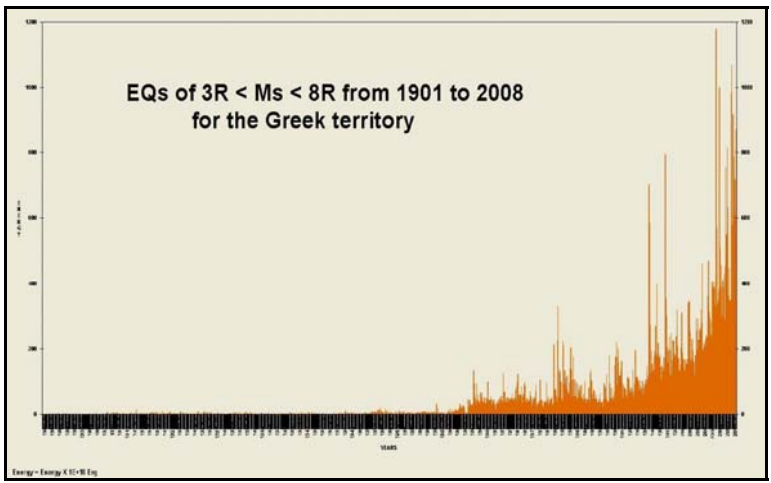

Fig. 21. EQs in Greece are shown as a function of time for the time period of 1901 to 2008 with a sampling interval of 1 month.

One could argue the agreement of figure **(20)** to the one of **(21)** on the ground that the more EQs are recorded the more energy release must be observed. Actually, the figure **(21)** represents mostly the increase of the microseismicity in the Greek territory, which in turn indirectly suggests large seismicity in the future. A more detailed view of graphs **(20)** and **(21)** are presented in figures **(22)** and **(23)**. These graphs have been calculated for the time period from 1960 to 2008.

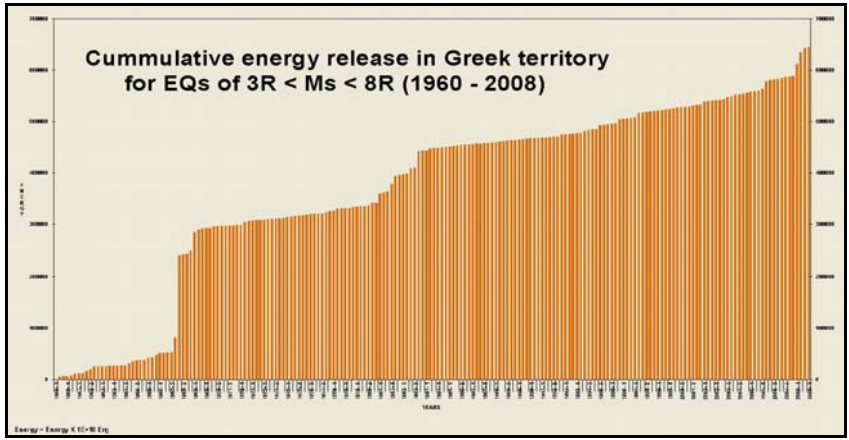

Fig. 22. Cumulative seismic energy release is shown of the Greek territory for the time period of 1960 to 2008. Sampling interval = 3 months.

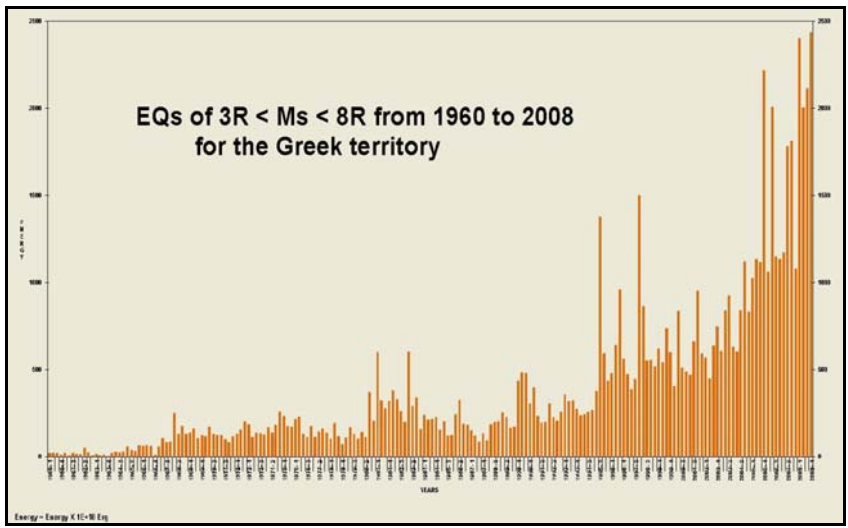

Fig. 23. EQs in Greece are shown as a function of time for the time period of 1960 to 2008 with sampling interval of 3 months.

Finally, a shorter period of time is selected in order to determine the accelerating deformation status of the Greek territory during the last **23** years. The cumulative seismic energy release is shown in next figure **(24)**. It is evident that accelerating deformation has been activated for the entire Greek seismogenic area.

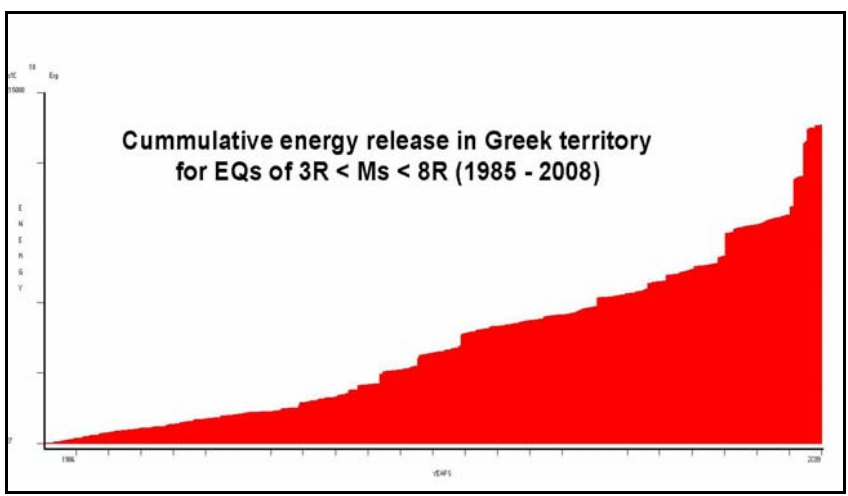

Fig. 24. Cumulative seismic energy release calculated for the time period from 1985 to 2008.

The data of figure **(24)** were fitted with a **6$^{th}$** degree polynomial in order to smooth the seismic energy release data. In the following figure **(25)** the entire operation is presented along with the time of occurrence of the Andravida EQ which is indicated by a red arrow.

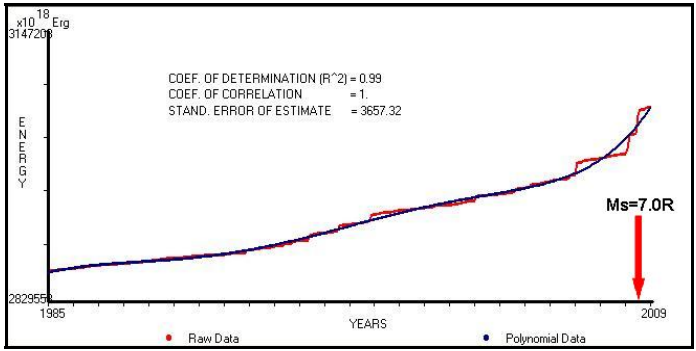

Fig. 25. Cumulative seismic energy release data for the time period from 1985 to 2008 (red line) fitted with a **6th** degree polynomial (black line). The time of occurrence of the Andravida EQ is indicated by a red arrow.

It is evident that the Andravida EQ did occur at the drastically increasing (accelerating) part of the graph of the seismic energy release. The accelerating deformation had increased rapidly during the last **3** years (2006 – 2008). Therefore, it is justified to say that due to this increase of energy release a large EQ could take place shortly (in terms of a couple of years). A sharper picture of the accelerating deformation can be obtained when the gradient of the cumulative seismic energy release is calculated from the fitted polynomial. The latter is shown in the following figure **(26)**.

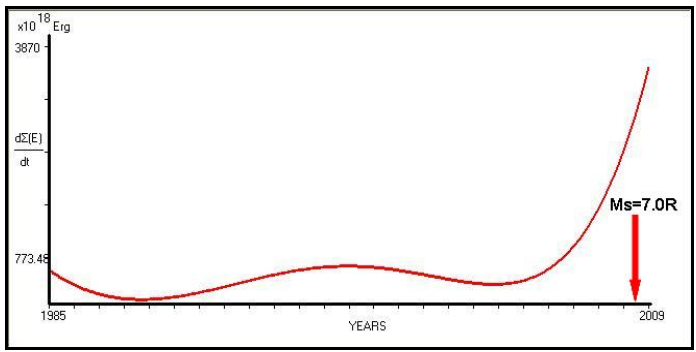

Fig. 26. Gradient (red line) of the cumulative seismic energy release calculated analytically from the 6th degree polynomial fitted to the cumulative seismic energy release data of figure **(25)**. The red arrow indicates the time of occurrence of the Andravida EQ.

It is evident that the gradient calculation resolved better the change in time of the acceleration deformation due to its high-pass filtering properties.

### 2.5.2. Andravida regional seismogenic Area.

After having observed the acceleration deformation status of the Greek territory and since the epicentral area of the pending EQ has already been calculated **(fig. 10)** it is interesting to investigate the accelerating deformation status of the regional seismogenic area of Andravida EQ. Consequently, a wider area around the calculated epicenter of the pending EQ has been selected for which the seismic energy release will be determined. The selection of the study area was made in such a way so that it included the main tectonic fracture zones and faults present in it. The study area is presented in the following figure **(27)**. The Andravida EQ regional seismogenic area is identified by a blue trapezoid while the main tectonic features of the Greek territory are represented by brown thick lines (Thanassoulas, 1998).

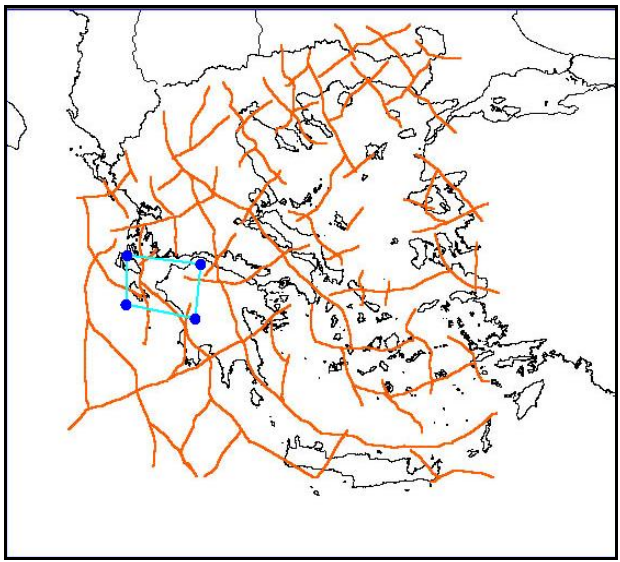

Fig. 27. Assumed Andravida EQ regional seismogenic area (blue frame). Brown thick lines represent the deep fracture zones – faults of the lithosphere.

The calculated cumulative seismic energy release through the Andravida EQ seismogenic regiona area is shown in the following figure **(28)**. This graph represents the cumulative energy release for the time period from 1961 to 2008.

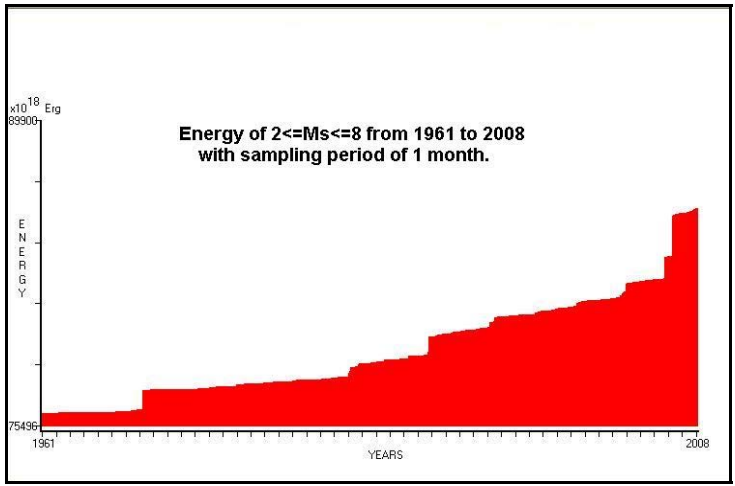

Fig. 28. Cummulative seismic energy release through the assumed seismogenic area for the period of 1961 - 2008 (just before the Andravida EQ).

The data of figure **(28)** have been smoothed by fitting a **6th** degree polynomial as it is shown in the following figure **(29)**. In this case too, as it was observed for the entire Greek territory, increasing accelerating deformation is observed for the last five years.

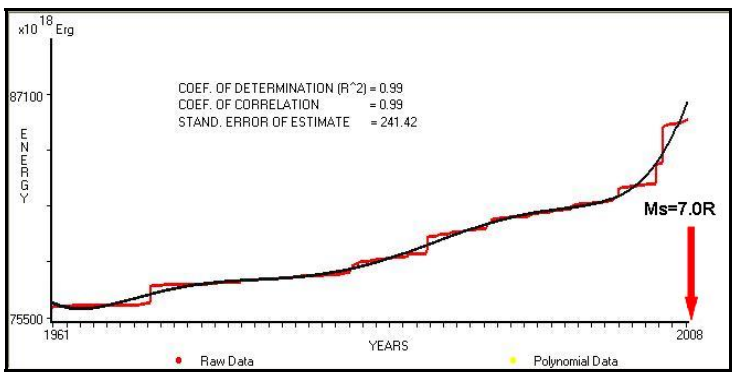

Fig. 29. Cumulative seismic energy release data for the time period from 1961 to 2008 (red line) fitted with a **6th** degree polynomial (black line). The time of occurrence of the Andravida EQ is indicated by a red arrow.

Again a sharper picture of the accelerating deformation can be obtained when the gradient of the cumulative seismic energy release is calculated from the fitted polynomial. The latter is shown in the following figure **(30)**.

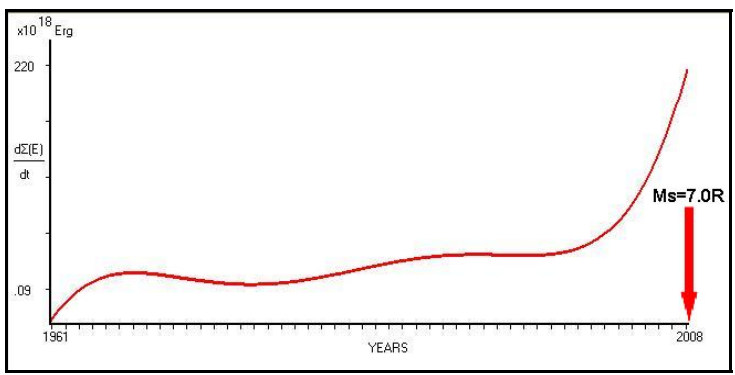

Fig. 30. Gradient (red line) of the cumulative seismic energy release calculated analytically from the **6th** degree polynomial fitted to the cumulative seismic energy release data of figure **(29)**. The red arrow indicates the time of occurrence of the Andravida EQ.

In this case the cumulative seismic energy release starts to increase for almost **7** years before the Andravida seismic event.

So far, it has been demonstrated that the Andravida EQ had affected, in terms of accelerating deformation, not only the regional Andravida seismogenic area but the entire Greek territory too due to its large magnitude.

Next, by manipulating the seismic energy release it is possible to calculate the expected maximum magnitude of the pending earthquake by applying the "lithosperic seismic energy flow model" (Thanassoulas, 2008). The entire methodology is presented in the following figure **(31)**. The calculated expected magnitude is: **Ms = 7.29R** compared to **Ms = 7.0R** obtained by seismological methods.

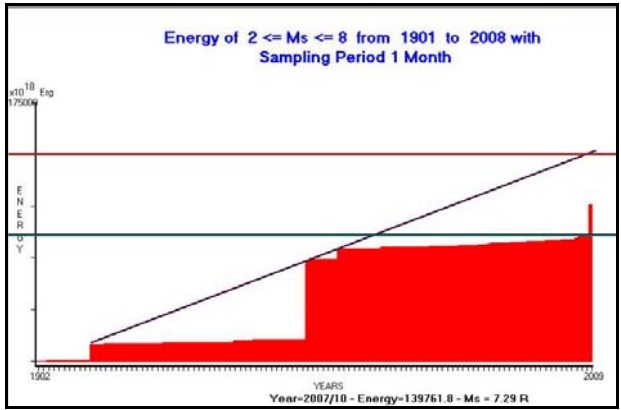

Fig. 31. Magnitude determination of the Andravida EQ. The magnitude calculated by seismological **(NOA)** methods is **Ms = 7.0R** compared to **Ms = 7.29R** calculated by the application of the "Lithospheric Seismic Energy Flow Model".

An obvious question that is raised immediately is whether the Andravida seismogenic area has been seismically discharged after the 2008 large seismic event or in the opposite case what is a probable magnitude of a future EQ. In order to answer this question the expected future magnitude has been calculated by taking as lower reference stored seismic energy level the one indicated by the Andravida EQ seismic energy release. The rate of seismic charge of the regional seismogenic area is the same (inclined line) as in figure **(31)**. Observe the use of the seismic energy reference level at the far right part (spike) in both graphs of figures **(31, 32)**.

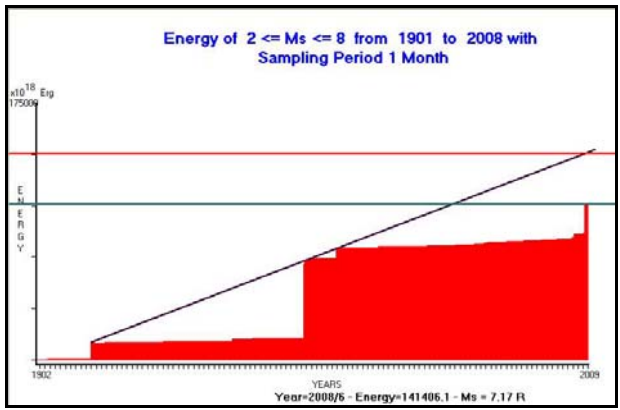

Fig. 32. Calculation of the expected magnitude for a future EQ in the Andravida regional seismogenic area. The seismic energy that is still stored in that area is capable to be released through a future EQ of **Ms = 7.17R**.

The figure **(32)** suggests that although a large EQ did occur in the Andravida seismogenic area this by no means indicates that this area has been seismically discharged. After all, for releasing the energy of an **Ms=7.0R** earthquake in terms of earthquakes of **Ms=6.0R** it takes about **33** EQs of such magnitude which actually was not (in terms of aftershocks) the case of the Andravida EQ.

Finally, the Andravida EQ location is compared to the seismic potential maps of Greece calculated for the years 2000 and 2005 (Thanassoulas, 2007; Thanassoulas et al. 2010). This is demonstrated in the following figure **(33)**.

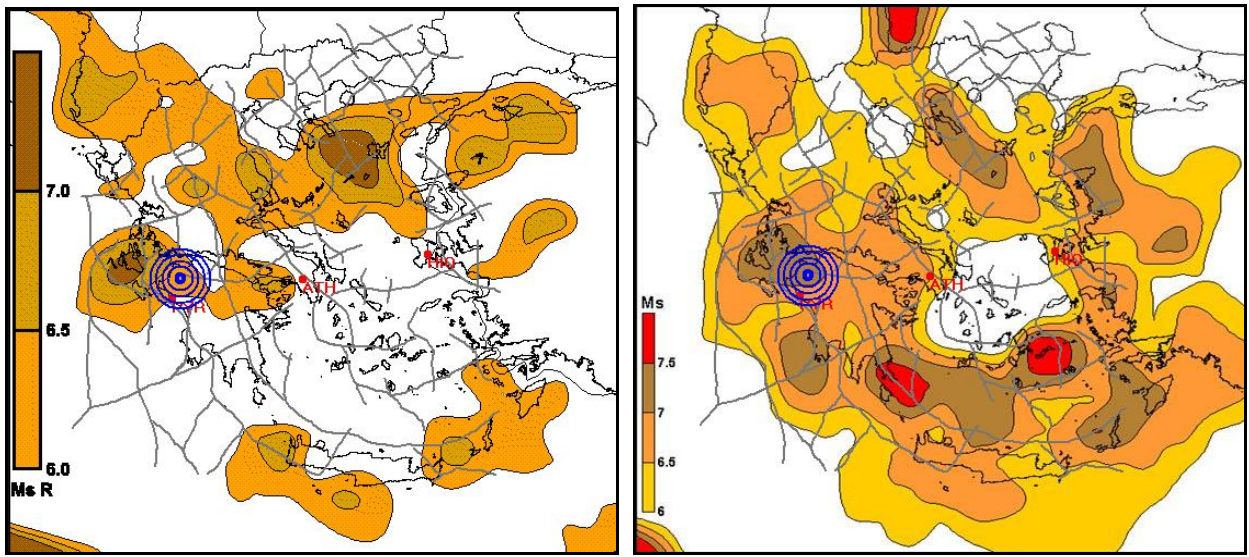

Fig. 33. Andravida EQ (blue concentric circles) location in relation to the seismic potential map of Greece calculated for the years 2000 (left) and 2005 (right).

In figure **(33)** left, the Andravida EQ is located in an area which is characterized by a seismic potential equivalent to an earthquake of magnitude **6.0R < Ms < 6.5R**. By that time no accelerating deformation had been observed in the same seismogenic area as it was shown by the figures **(29, 30)**. Five years after, on 2005, the seismic potential has drastically increased. The Andravida EQ is located in an area characterized by a seismic potential equivalent to an earthquake of magnitude **7.0R < Ms < 7.5R**. Three years after, on 2008, the Andravida EQ did occur with a magnitude of **Ms = 7.0R** according to the seismological methods of **NOA** or **Ms = 7.29** according to the "lithospheric seismic energy flow model".

### 3. Conclusions.

Summarizing all the afore presented results of the "a posteriori" analysis of the data regarding the Andravida EQ it is concluded that the specific EQ could be predicted quite accurately in terms of "short-term" prediction. In detail:

- The analysis of the past seismicity of the Greek territory revealed that the entire Greek seismogenic area had been set at an accelerating deformation mode for at least the last 3 ÷ 4 years. Quite similar seismic charge conditions were met at the Andravida seismogenic area, after its identification by the analysis of the Earth's electric field recorded by **ATH** and **HIO** monitoring sites. Therefore, the Andravida EQ can be considered as predictable in terms of "medium-term" prediction. Furthermore, the Andravida EQ verified the yearly lithospheric deformation model due to the year's period tidal component since it did occur almost on the Aphelion of the Earth's orbit around the Sun.

- A shorter (a couple of days and at specific times in each day) time window for the occurrence of the Andravida EQ could be defined by the use of the tidal waves **K1, M1 (T=1, 14 days)** that trigger the lithospheric oscillation and the "strange attractor like" seismic electric precursor.

- Consequently, there was a progressively shortening of the time prediction window from medium-term to short-term and even to immediate-term by combining: past seismic energy release, tidal waves and seismic precursory signals ("strange attractor like").

- The epicenter of the Andravida EQ was determined, by triangulation, quite accurately by the analysis of the recorded Earth's electric field by **PYR** and **HIO** monitoring sites. The actual seismic precursory signals were obtained after applying the "noise injection" methodology. These calculations are based on the assumption of a homogeneous Earth's ground model which is valid for the very long wavelength electric preseismic signals and the concept of the apparent point current source (Thanassoulas, 1991).

- Finally, the magnitude of the Andravida EQ was determined by the application of the "lithospheric seismic energy flow model" based on the energy conservation laws of physics. The latter complies quite well too, with the observed seismic potential (year 2005) of the regional Andravida area.

- In conclusion, all three seismic parameters (time, location and magnitude) which characterize a seismic event (evidently it refers to the large ones) were "a posteriori" successfully calculated for the Andravida EQ. The purpose of this work is to show that most of large EQs can be predicted quite accurately provided that the right methodologies (monitoring of appropriate physical parameters, use of physical models, mathematical analysis) and involvement of skilled scientific personnel and funds are available. The presented methodology can be applied "a priori" provided that the appropriate establishment (monitoring network, scientific personnel, and funds) exists.

- All the original data used in this work can be downloaded from the www.earthquakeprediction.gr internet site.

### 4. References.

Rudman, J. A., Ziegler, R., and Blakely, R., 1977. Fortran Program for Generation of Earth Tide Gravity Values, Dept. Nat. Res. State of Indiana, Geological Survey.
Sarlis, V.N., Skordas, S.E., Lazaridou, S.M., Varotsos, A.P. 2008. Investigation of the seismicity after the initiation of a Seismic Electric Signal activity until the main shock., arxiv:0802.3329 v4 [cond-mat.stat-mech]
Thanassoulas, C., 1991. Determination of the epicentral area of three earthquakes (Ms>6) in Greece, based on electrotelluric currents recorded by the VAN network., Acta Geoph. Polonica, Vol. XXXIX, no. 4, pp. 273 – 287.
Thanassoulas, C., 1998. Location of the seismically active zones of the Greek area by the study of the corresponding gravity field., Open File Report: A.3810, Inst. Geol. Min. Exploration (IGME), Athens, Greece, pp. 1-19.
Thanassoulas, C., 2007. Short-term Earthquake Prediction, H. Dounias & Co, Athens, Greece. ISBN No: 978-960-930268-5
Thanassoulas, C. 2008. The seismogenic area in the lithosphere considered as an "Open Physical System". Its implications on some seismological aspects. Part – II. Maximum expected magnitude. arxiv:0807.0897 v1 [physics.geo-ph]
Thanassoulas, C., Klentos, V., Verveniotis, G. 2008. Extracting preseismic electric signals from noisy Earth's electric field data recordings. The "noise injection" method. arXiv:0807.4298v1 [physics.geo-ph].
Thanassoulas, C., Klentos, V., Verveniotis, G. 2008a. On a preseismic electric field "strange attractor" like precursor observed short before large earthquakes. arXiv:0810.0242v1 [physics.geo-ph].
Thanassoulas, C., Klentos, V., Verveniotis, G., Zymaris, N. 2009. Seismic electric precursors observed prior to the 6.3R EQ of July 1st, 2009 Greece, and their use in short-term earthquake prediction. arXiv:0908.4186v1 [physics.geo-ph].
Thanassoulas, C., Klentos, V., Tsailas, P., Verveniotis, G., Zymaris, N. 2009a. Medium - long term earthquake prediction by the use of the oscillating electric field (T = 365 days) generated due to Earth's orbit around the Sun and due to its consequent oscillating lithospheric deformation. arXiv:0912.0818 v1 [physics.geo-ph].
Thanassoulas, C., Klentos, V. 2010. Seismic potential map of Greece calculated for the years 2005 and 2010. Its correlation to the large (Ms>=6.0R) seismic events of the 2000 - 2009 period. arXiv:1001.1491 v1 [physics.geo-ph].
Thanassoulas, C., Klentos, V. 2010a. How "Short" a "Short-term earthquake prediction" can be? A review of the case of Skyros Island,Greece, EQ (26/7/2001, Ms = 6.1 R). arXiv:1002.2162 v1 [physics.geo-ph].